\documentclass{aa}  
\usepackage{txfonts}
\bibpunct{(}{)}{;}{a}{}{,} % to follow the A&A style
\usepackage{amstext}

\begin{document}


\title{Comparison of the shape and temporal evolution of even and odd solar cycles}

\author{Jouni Takalo
\and
Kalevi Mursula
}

\institute{Space physics and astronomy research unit, University of Oulu,
POB 3000, FIN-90014, Oulu, Finland\\
\email{jouni.j.takalo@oulu.fi}
}

\date{Received: }

\abstract {}
{We study the difference in the shape of solar cycles for even and odd cycles using the Wolf sunspot numbers and group sunspot numbers of solar cycles 1-23. We furthermore analyse the data of sunspot area sizes for even and odd cycles SC12-SC23 and sunspot group data for even and odd cycles SC8-SC23 to compare the temporal evolution of even and odd cycles.}
{We applied the principal component analysis (PCA) to sunspot cycle data and studied the first two components, which describe the average cycle shape and cycle asymmetry. We used a distribution analysis to analyse the temporal evolution of the even and odd cycles and determined the skewness and kurtosis for even and odd cycles of sunspot group data.}
{The PCA confirms the existence of the Gnevyshev gap (GG) for solar cycles at about 40\% from the start of the cycle. The temporal evolution of sunspot area data for even cycles shows that the GG exists at least at the 95\% confidence level for all sizes of sunspots. On the other hand, the GG is shorter and statistically insignificant for the odd cycles of aerial sunspot data. Furthermore, the analysis of sunspot area sizes for even and odd cycles of SC12-SC23 shows that the greatest difference is at 4.2-4.6 years, where even cycles have a far smaller total area than odd cycles. The average area of the individual sunspots of even cycles is also smaller in this interval.
The statistical analysis of the temporal evolution shows that northern sunspot groups maximise earlier than southern groups for even cycles, but are concurrent for odd cycles. Furthermore, the temporal distributions of odd cycles are slightly more leptokurtic than distributions of even cycles. The skewnesses are 0.37 and 0.49 and the kurtoses 2.79 and 2.94 for even and odd cycles, respectively. The correlation coefficient between skewness and kurtosis for even cycles is 0.69, and for odd cycles, it is 0.90.}
{The separate PCAs for even and odd sunspot cycles show that odd cycles are more inhomogeneous than even cycles, especially in GSN data. Even cycles, however, have two anomalous cycles: SC4 and SC6. The  variation in the shape of the early sunspot cycles suggests that there are too few and/or inaccurate measurements before SC8.
According to the analysis of the sunspot area size data, the GG is  more distinct in even than odd cycles. This may be partly due to sunspot groups maximizing earlier in the northern than in the southern hemisphere for even cycles. We also present another Waldmeier-type rule, that is, we find a correlation between skewness and kurtosis of the sunspot group cycles.}


\titlerunning{Comparison of even and odd solar cycles}
\authorrunning{Takalo and Mursula}

\maketitle

\section{Introduction}

Almost two hundred years ago, it was noted that the occurrence of sunspots is cyclic. However, there are differences in the cycles; for instance, the length of the cycle changes from 9.0 to 13.7 years and the shape of the cycle changes somewhat between cycles and also between hemispheres. \cite{Waldmeier_1935} found that each cycle  is also asymmetric such that the ascending phase is  shorter than the declining phase, and that there is anti-correlation between cycle amplitude and the length of the ascending phase of the cycle \citep{Waldmeier_1939}.

\cite{Gnevyshev_1967} suggested that the solar cycle is characterised by two periods of activity, and these lead to a double peak with the so-called Gnevyshev gap (GG) in between \citep{Gnevyshev_1977}. \cite{Feminella_1997} studied the long-term behaviour of several solar activity parameters and found that maxima occur at least twice: first, near the end of the rising phase, and then in the early years of the declining phase. \cite{Norton_2010} analysed the sunspot cycle double peak and the \emph{GG} between them to determine if the double peak is caused by averaging of two hemispheres that are out of phase \citep{Temmer_2006}. They confirmed previous findings, however, that the GG is a phenomenon that occurs in the separate hemispheres and is not due to a superposition of sunspot indices from hemispheres that are slightly out of phase.

Most of the even-odd cycle comparisons have concentrated on the mutual strength of preceding cycles. These are referred to as the so-called Gnevyshev-Ohl rule, which is an expression of the general 22-year variation of cycle amplitudes and intensities, according to which even cycles are on average about 10\%-15\% lower than following odd cycles \citep{Mursula_2001}. There have been some violations of this rule, however; the last occurred between the cycle pair SC22-SC23 \citep{Javaraiah_2012, Javaraiah_2016}.

Another common subject has been the north-south asymmetry in solar sunspots and other activity (see some of the recent publications \citep{Carbonell_2007, Li_2009, Chang_2012, Hathaway_2015, Javaraiah_2016, Vernova_2016, Badalyan_2017, Chowdhury_2019}). Many studies have also been conducted of the spatial (latitudinal) distribution of sunspots and their migration throughout the solar cycle \citep{Ivanov_2011, Chang_2012, Jiang_2011, Munoz-Jaramillo_2015, Santos_2015, Leussu_2016, Leussu_2017, Mandal_2017, Zhang_2018}. Less attention has been paid to the temporal distribution of the total strength of the sunspots, sunspot groups, and areas throughout the solar cycle (except for the indices themselves). 
\cite{Leussu_2016} in particular studied the latitude evolution and the timing of the sunspot groups in butterfly wings by characterising three different categories: the latitude at which the first sunspot groups appear, the maximum latitude of the sunspot group occurrence in each wing, and the latitude at which the last sunspot group appears. The authors derived several statistical measures based on these variables.
Some studies have investigated the distribution of the accumulated area or number of sunspots as a function of area size \citep{Zharkov_2005, Santos_2015}

In this study we use the principal component analysis (PCA) to calculate the average shape of the sunspot cycles separately for even and odd cycles using the SSN and GSN of sunspot cycles 1-23 and sunspot areas for cycles SC12-SC23. Furthermore, we study the temporal evolution of sunspot areas for even and odd cycles of SC12-SC23 and the temporal distribution for sunspot group data for cycles SC8-SC12. This paper is organised as follows: Section 2 presents the data and methods. In Section 3 we present the results of the PCA for the cycle shape using sunspot numbers and group sunspot numbers for even and odd cycles. In Section 4 we analyse the sunspots area sizes for even and odd cycles using the PCA. Section 5 presents the temporal analysis of sunspot areas and sunspot groups for even and odd solar cycles. We give our conclusions in Section 6.

\section{Data and methods}

\subsection{Sunspot indices}

\begin{figure*}
        \centering
        \includegraphics[width=0.9\textwidth]{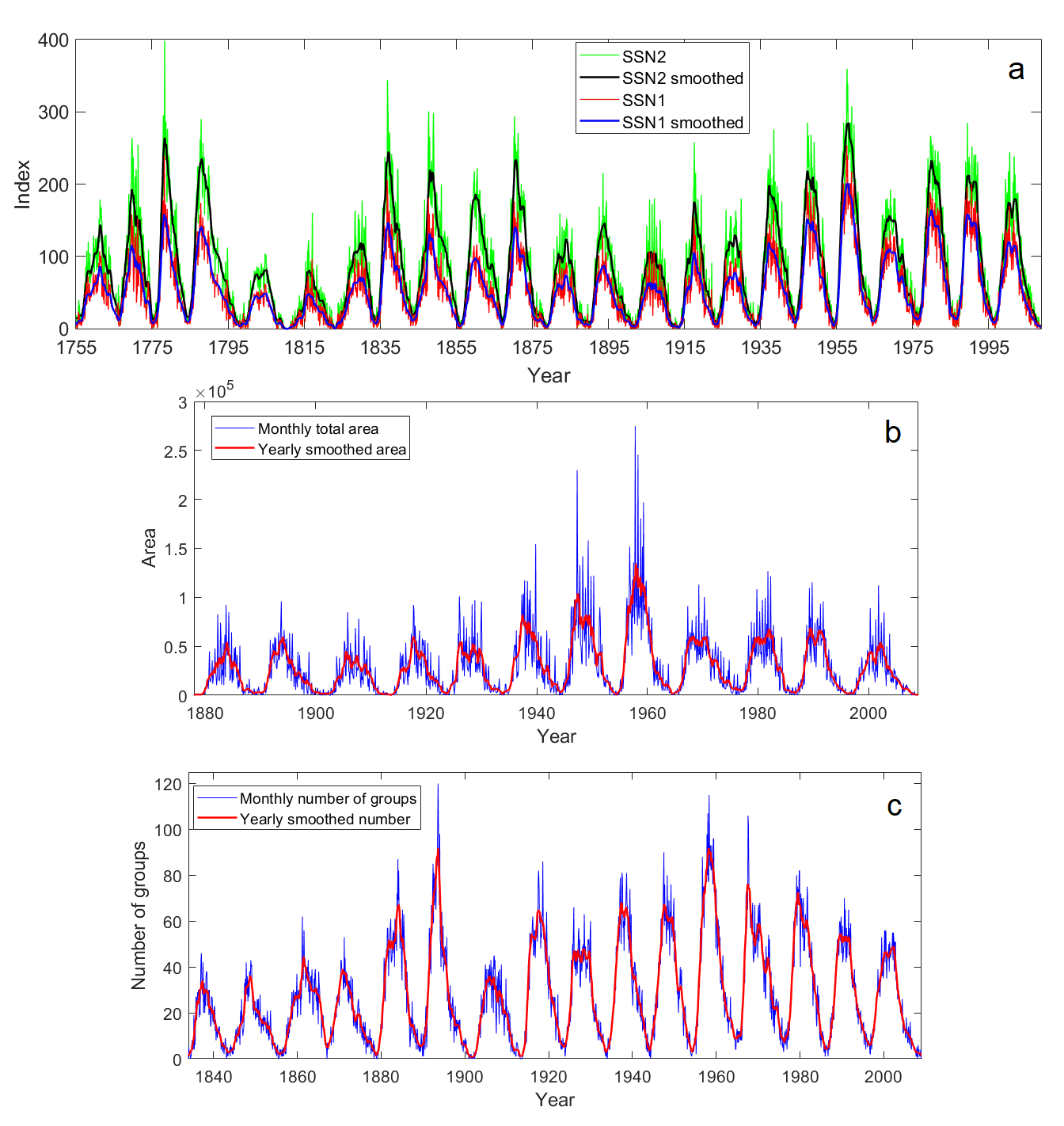}
                \caption{a) Sunspot indices, SSN1 and SSN2, for the cycles SC1-SC23 and their Gleissberg-smoothed indices. b) Sunspot area index and its yearly smoothed index for the cycles SC12-SC23. c) Number of sunspot groups and their yearly smoothed number for cycles SC8-SC23.}
                \label{fig:Solar_indices}
\end{figure*}

Because the first complete sunspot cycle included in the SSN started in March 1755, it was numbered SC1 by Rudolf Wolf. This numbering of sunspot cycles is still in use. The initial sunspot number series (here called SSN1) was reconstructed at the Z\"urich  Observatory until 1980, and at the Royal Observatory of Belgium since 1981. Following the change in reconstruction method in 1981, the current version of the SSN series is called the international sunspot number (ISN). The ISN series was recently modified to a version 2.0 that is supposed to present a preliminary correction of the past inhomogeneities in the SSN1 series \citep{Clette_2014}.
Figure \ref{fig:Solar_indices}a shows both sunspot indices (SSN1 and SSN2) for the cycles SC1-SC23 and their  Gleissberg-smoothed (box-car smoothing over 13 months such that the end points have half the weight of the other points) indices. The new index 2.0 gives higher peaks than the old index for the whole interval 1955-2009, but the shape of the cycles is very similar. This is especially well seen in the smoothed indices. In this study we use monthly indices of SSN1, but we verified that using SSN2 gives very similar results. The dates of the sunspot minima and the cycle lengths for SSN1 are shown in Table1.

\subsection{GSN index}

\begin{table}
\begin{center}
\caption{Sunspot cycle lengths (in years) and dates (fractional years, and year and month) of (starting) sunspot minima for SSN1 \citep{NGDC_2013}.}
\begin{tabular}{ c  c  c  c }
\hline
  Sunspot cycle &SSN1 fractional  &Year and month     &Cycle length  \\
      number    &year of minimum   & of SSN1 min     &    (years) \\
        \hline 
1     &1755.2   &1775 March  & 11.3  \\
2     &1766.5   &1766 June  & 9.0  \\
3     &1775.5   &1775 June  & 9.2  \\
4     &1784.7   &1784 September  &     13.7  \\
5     &1798.4   &1798 May  & 12.2  \\
6     &1810.6   &1810 August  & 12.7  \\
7     &1823.3   &1823 April & 10.6  \\
8     &1833.9   &1833 November & 9.6  \\
9     & 1843.5  &1843 July  & 12.5  \\
10    & 1856.0  &1855 December  & 11.2  \\
11    & 1867.2  &1867 March  & 11.8  \\
12    & 1879.0  &1878 December  & 10.6  \\
13    & 1889.6  &1889 August  & 12.1  \\
14    & 1901.7  &1901 September  & 11.8  \\
15    & 1913.5  &1913 July  & 10.1  \\
16    & 1923.6  &1923 August & 10.1  \\
17    & 1933.7  &1933 September  & 10.4  \\
18    & 1944.1  &1944 February  & 10.2  \\
19    & 1954.3  &1954 April  & 10.5  \\
20    & 1964.8  &1964 October  & 11.7  \\
21    & 1976.5  &1976 June & 10.2  \\
22    & 1986.7  &1986 September  & 10.1  \\
23    & 1996.8  &1996 October  & 12.2  \\
24    & 2009.0  &2008 December  &       \\
\hline
\end{tabular}
\end{center}
\end{table}

Although the group sunspot number (GSN) index starts as early as 1610, see \cite{Hoyt_1998}, its coverage is scarce until solar cycle 1 (SC1), and some monthly values up to SC5 are still missing. The GSN series also ends in 1995, that is, SC23 is missing. We therefore filled in the gaps in the monthly GSN data in SC1-SC4 using linear interpolation. In order to continue the GSN after SC22, we used the recently published GSN time series \citep{Chatzistergos_2017} and adjusted it to the level of the average GSN time series in SC15-SC22 using SSN1 as reference index. It seems that the minima of the GSN index are not always the same as in SSN1. Therefore we defined the minima of the GSN data using GSN time series \citep{Takalo_2018}. The dates of GSN minima and their difference to SSN1 minima are shown in Table 2.

\begin{table}
\begin{center}
\small
\caption{Dates (fractional years, and year and month) of (starting) minima of GSN cycles, GSN cycle lengths, and their difference to SSN1 minima (in months).}
\begin{tabular}{  c  c  c  c  c }
\hline
    Cycle &  Fractional year  &Year and month  &Cycle  &Diff. to \\
    number      &of minimum    & of minimum    &length & SSN1 min\\
        \hline 
1      &1755.2   &1775 March & 11.2    & 0 \\
2      &1766.4   &1766 May & 9.1  & -1 \\
3      &1775.5   &1775 June & 9.0  & 0 \\ 
4      &1784.5   &1784 July & 14.2  & -2 \\
5      &1798.7   &1798 September & 11.8 & +4  \\
6      &1810.5   &1810 July & 12.8 & -1 \\
7      &1823.3   &1823 April & 10.6 &  0  \\
8      &1833.9   &1833 November & 9.7 &  0 \\
9      & 1843.6  &1843 August & 12.5 & +1  \\
10     & 1856.1  &1856 February  & 11.2 & +2  \\
11     & 1867.3  &1867 April  & 11.7 & +1  \\
12     & 1879.0  &1879 January & 10.9  &  +1 \\
13     & 1889.9  &1889 November & 12.1 & +3  \\
14     & 1902.0  &1901 December & 11.6  & +3 \\
15     & 1913.6  &1913 August & 10.2  & +1  \\
16     & 1923.8  &1923 October & 10.1 & +2 \\
17     & 1933.9  &1933 November & 10.4  & +2 \\
18     & 1944.3  &1944 April & 10.0 & +2 \\
19     & 1954.3  &1954 April & 10.4 &  0  \\
20     & 1964.7  &1964 September & 11.8  & -1  \\
21     & 1976.5  &1976 June & 10.0 &  0 \\
22     & 1986.5  &1986 June & 10.2 & -3  \\
23     & 1996.7  &1996 September & 12.4 & -1  \\
24     & 2009.1  &2009 February  &   &  +2   \\

\hline
\end{tabular}
\end{center}
\end{table}

\subsection{Temporal sunspot area data}

In the sunspot area analysis we used the database of the Royal Observatory, Greenwich-USAF/NOAA Sunspot Data \citep{Nasa_MSFC_2017} for 1874-2016. This database contains among others the time, latitude, and area size (in millionths of solar hemisphere, MH) for individual sunspots for cycles SC12-SC23. Here we used the total (corrected) area consisting of both the umbral (darker) and penumbral (lighter border area) regions. The minima are same as in the SSN analysis, starting from December 1878 (1878.9 in decimal year). Figure \ref{fig:Solar_indices}b shows the sunspot area index (here the unit is 0.1 years) and its yearly (10 points) smoothed index. It is evident that the area data are different from the sunspot index. For example, the total areas of cycles 12-16 are almost similar, while there are differences in the heights of the sunspot number index. The reason is that the sunspot number is calculated from sunspot groups and individual spots, regardless of their size. Furthermore, \cite{Takalo_2020} has shown that large sunspots occur mainly at latitudes 10-25, except for a gap (the GG) at about 15 degrees, while smaller sunspots tend to be located at lower latitudes on average. As a consequence, large sunspots are lacking at the start and near the end of the sunspot cycle.

\subsection{Temporal sunspot group data}

\begin{figure*}
        \centering
        \includegraphics[width=0.95\textwidth]{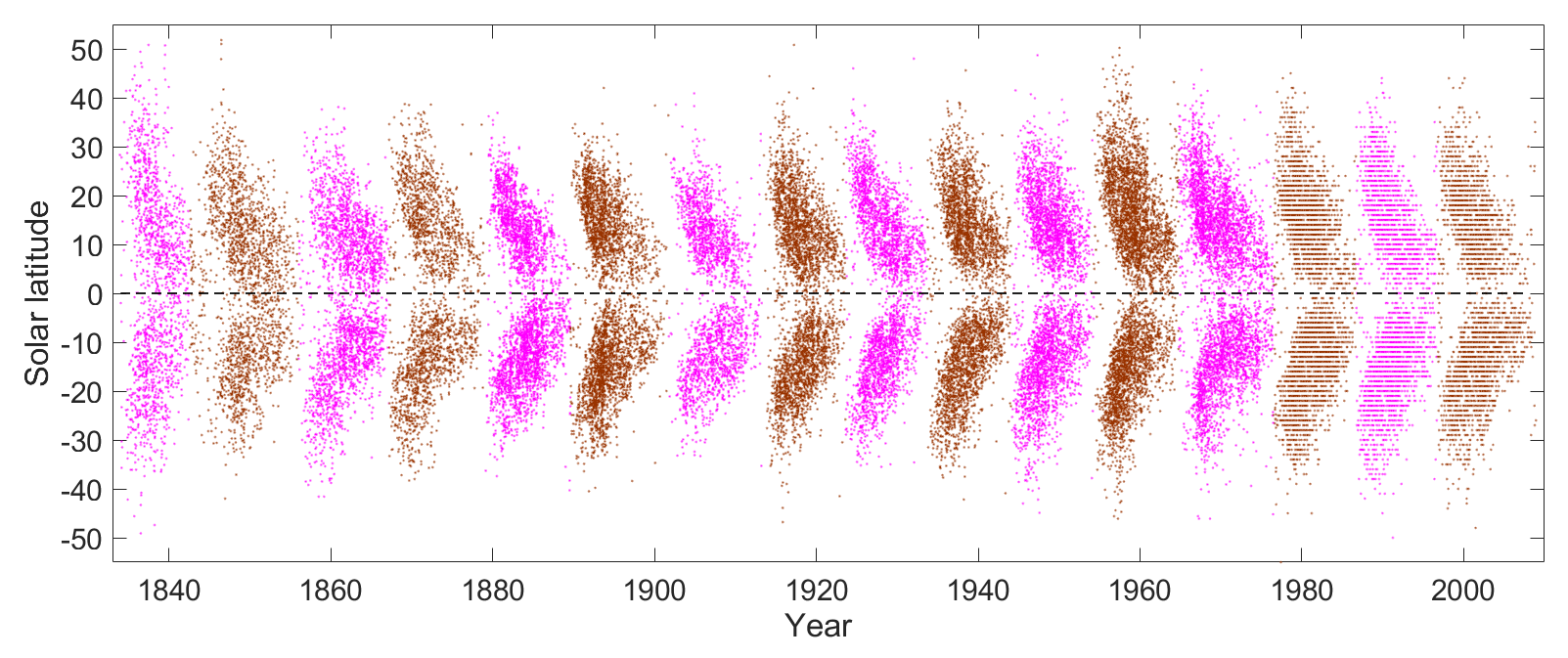}
                \caption{Butterfly pattern of the sunspot groups. The vertical lines are the corresponding cycle maxima (adopted from the National Geophysical Data Center (NGDC), Boulder, Colorado, USA (ftp.ngdc.noaa.gov).}
                \label{fig:Leussu_data}
\end{figure*}

In the group data analysis we used the data set of sunspot groups in the southern and northern wings for cycles SC8-SC23 by Leussu et al. \citep{Leussu_2017}. These data include the time and latitude for sunspot groups for cycles SC8-SC23 and is shown as the butterfly pattern in Fig. \ref{fig:Leussu_data}. Figure \ref{fig:Solar_indices}c shows the same data as an index (unit 0.1 years) and its yearly smoothed index.

\subsection{PCA method}

The PCA is a technique for reducing the dimensionality of data sets, that is,  increasing interpret-ability, but at the same time minimising information loss. For a large number of correlated variables, the PCA finds combinations of a few uncorrelated variables that describe the majority of the variability in the data. The first principal component (PC1) carries most of the variance and therefore describes the main feature of the whole data set. The second principal component (PC2) is perpendicular to PC1 and accounts for second largest part of the variance. The third principal component (PC3) is perpendicular to both PC1 and PC2 and is usually less significant \citep{Jolliffe_2002, Jolliffe_2016}.

In our case, the two main components, PC1 and PC2, are enough to describe the shape of the the solar cycles because they account for 80-90\% of the whole variance in the data (except for the sunspot area analysis, where the data are more heterogeneous). The PC1 gives the average shape of the solar cycle, and PC2 is the leading correction component compared to the average shape. The higher PCs usually describe some anomalous features that are present only in some  cycles of the data set. Because the PCA is a matrix-based method, sunspot cycles need to have equal length. To this end, we resampled the monthly sunspot values so that all cycles had the same length of 133 time steps (months). Before applying the PCA to the resampled sunspot cycles, we standardised each individual cycle to have zero mean and unit standard deviation. This guarantees that all cycles have the same weight in the study of their common shape (see \citep{Takalo_2018} and the appendix for a more detailed description of the method).

\subsection{Statistical methods}

\subsubsection{Generalised extreme value distribution}

The probability density function (PDF) of the generalised extreme value (GEV) distribution is expressed as
\begin{equation}
f_{S}\left(s;k\right) = \left(1+ks\right)^{(-1/k)-1)}\,e^{-(1+ks)^{-1/k}},\;k\neq0  ,
\end{equation}
and
\begin{equation}
f_{S}\left(s;0\right) = e^{-s}\,exp\left[{-e^{-s}}\right],\;k=0
,\end{equation}
where s is the standardised variable  $s=(x-\mu)/\sigma$. Here $\mu$ and $\sigma$ are the location and scale parameters, respectively, and k is the shape parameter. It is clear that this expression follows from the definition of the cumulative distribution function (CDF) $F=e^{-(1+ks)^{-1/k}},\;k\neq0 $. If k equals zero, the probability function is defined separately, but in our case, k$\neq$0 is always valid. An interesting fact of the GEV distribution is that if we have N data sets from the same distribution and we create a new data set that includes the extreme values from these N data sets, the resulting data set can be described by the GEV distribution \citep{Kotz_2000, Coles_2001}.

\subsubsection{Negative log-likelihood}

The likelihood function $L\left(\theta\right)$ is defined as
\begin{equation}
        L\left(\theta\right) = \prod^{n}_{i=1}f_{\theta}\left(x_{i}\right) ,
\end{equation}

if each variable $x_{i}$ is independent and from the same distribution $f_{\theta}$. The set of parameters $\theta$ of the distribution, which maximises $L\left(\theta\right)$ is called a maximum likelihood estimator (MLE) and is denoted $\theta_{L}$. It is often easier to maximise the log-likelihood function, $log\,L\left(\theta\right)$, and because the (natural) logarithmic function increases monotonically, the same value maximises both $L\left(\theta\right)$ and $log\,L\left(\theta\right)$. Because the log-likelihoods are here always negative, we calculated the minimum value for the negative log-likelihood (NLogL) \citep{Forbes_2011}.

\subsubsection{Two-sample T-test}

The two-sample T-test for equal mean values is defined as follows: The null hypothesis assumes that the means of the samples are equal, that is, $\mu_{1}=\mu_{2}$. The alternative hypothesis is that $\mu_{1}\neq\mu_{2}$. The test statistic is calculated as
\begin{equation}
T = \frac{\mu_{1} - \mu_{2}}{\sqrt{{s^{2}_{1}}/N_{1} + {s^{2}_{2}}/N_{2}}} ,
\end{equation}

where $N_{1}$ and $N_{2}$ are the sample sizes, $\mu_{1}$ and $\mu_{1}$ are the sample means, and $s^{2}_{1}$ and $s^{2}_{2}$ are the sample variances. If the sample variances are assumed equal, the formula reduces to
\begin{equation}
T = \frac{\mu_{1} - \mu_{2}} {s_{p}\sqrt{1/N_{1} + 1/N_{2}}} ,
\end{equation}

where

\begin{equation}
s_{p}^{2} = \frac{(N_{1}-1){s^{2}_{1}} + (N_{2}-1){s^{2}_{2}}} {N_{1} + N_{2} - 2} .
\end{equation}

The rejection limit for the two-sided T-test is $\left|T\right| > t_{1-\alpha/2,\nu}$, where $\alpha$ denotes the significance level and $\nu$ the degrees of freedom. The values of $t_{1-\alpha/2,\nu}$ are published in T-distribution tables (\citep{Snedecor_1989, Krishnamoorthy_2006, Derrick_2016}).

\section{PCA of sunspot indices}
 
\begin{figure*}
        \centering
        \includegraphics[width=0.8\textwidth]{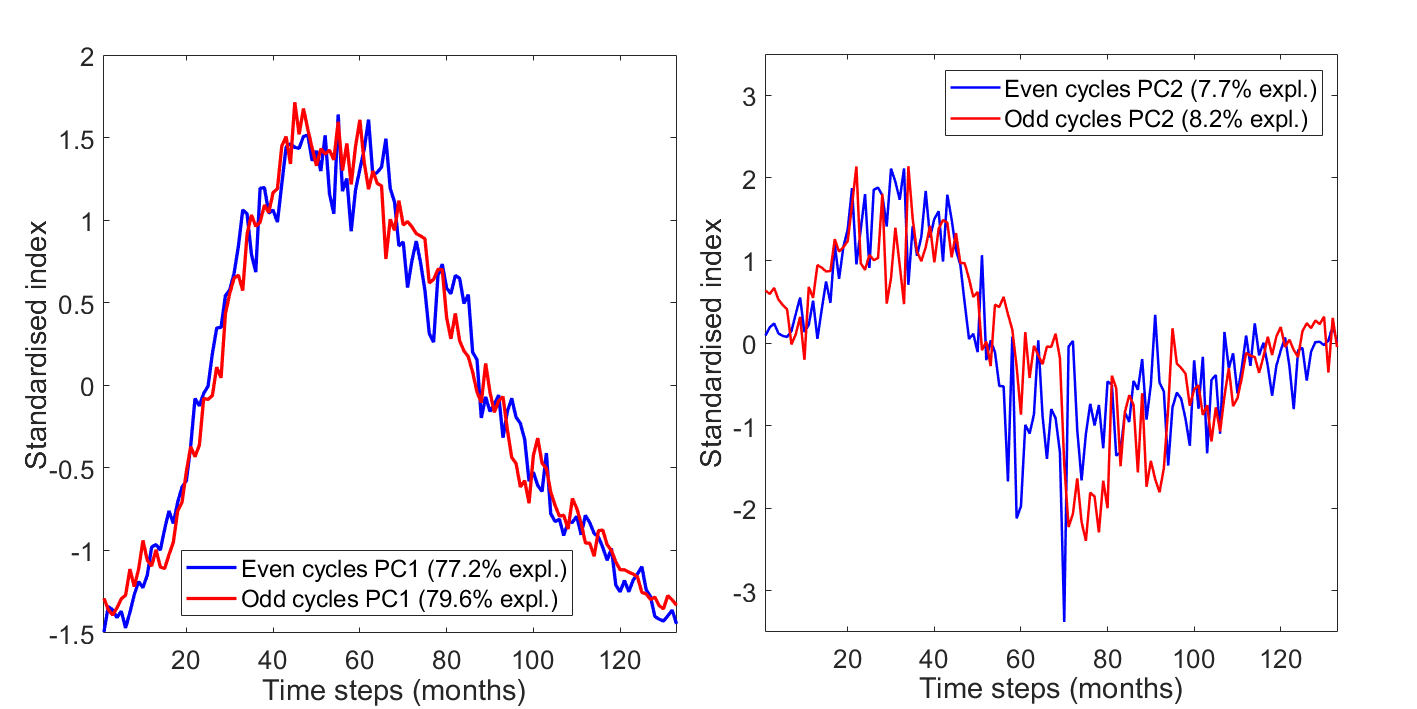}
                \caption{a) First and b) second principal components for the SSN1 even and odd solar cycles 1-23.}
                \label{fig:Even_odd_PCs}
\end{figure*}

We divided the cycles into two groups, even and odd numbered cycles between solar cycles 1-23. We then applied the PCA separately to these groups in order to study the differences between even and odd cycles. Figure \ref{fig:Even_odd_PCs} shows the first and second  principal components of even and odd solar cycles in panels \ref{fig:Even_odd_PCs}a and \ref{fig:Even_odd_PCs}b, respectively. The PC1s explain 77.2\% and 79.6\% and PC2s explain 7.7\% and 8.2\% of the total variance of the even and odd cycles, respectively. These two main PCs account for 84.9\% (even cycles) and 87.8\% (odd cycles) of the variation. It is evident that the first PCs are quite similar, while the PC2 differ more from each other. The correlation coefficients of the first PCs is 0.986 (p < $10^{-100}$), and the correlation coefficient of the PC2s  is 0.765 (p < $10^{-26}$).  PC1 has a gap after the maximum, the so-called Gnevyshev gap (GG) \citep{Gnevyshev_1967, Gnevyshev_1977, Storini_2003, Ahluwalia_2004, Bazilevskaya_2006, Norton_2010, Du_2015, Takalo_2018}, for both the even and odd cycles. They have a different form and place for odd and even cycle PC1s, however. Especially the gap for odd cycles is much narrower than the gap for even cycles. Another difference is that PC1 for even cycles has higher peaks in the declining phase of the cycle than PC1 for odd cycles.

\begin{figure}
        \centering
        \includegraphics[width=0.5\textwidth]{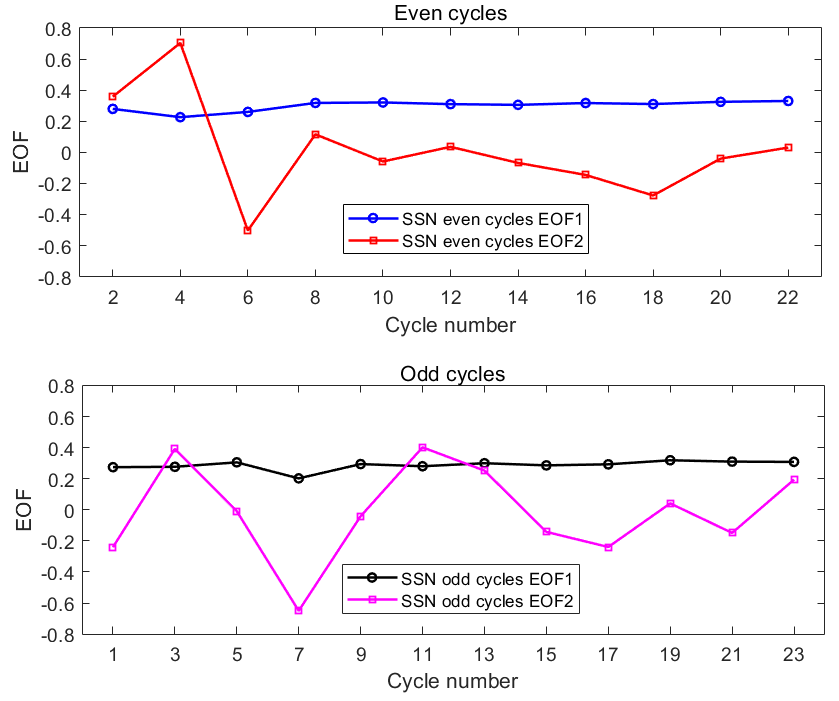}
                \caption{First two EOFs of a) even sunspot cycles and b) odd sunspot cycles.}
                \label{fig:EOFs_even_odd} 
\end{figure}

Figures \ref{fig:EOFs_even_odd}a and \ref{fig:EOFs_even_odd}b show the empirical orthogonal
functions (EOF) of the even and odd cycles, respectively. The EOF1s for odd cycles have almost equal weight for PC1, except for cycle 7. However, all cycles in the18th century, cycles 2, 4, and 6, have less weight than other cycles in the PC1 of even cycles. On the other hand, the EOFs of PC2 for odd cycles vary considerably between individual cycles, while the EOFs of PC2 for even cycles have less variation. In particular, after the 18th century, the EOFs of even cycles are very near zero, while the EOFs of odd cycles vary more strongly.

\begin{figure*}
        \centering
        \includegraphics[width=0.8\textwidth]{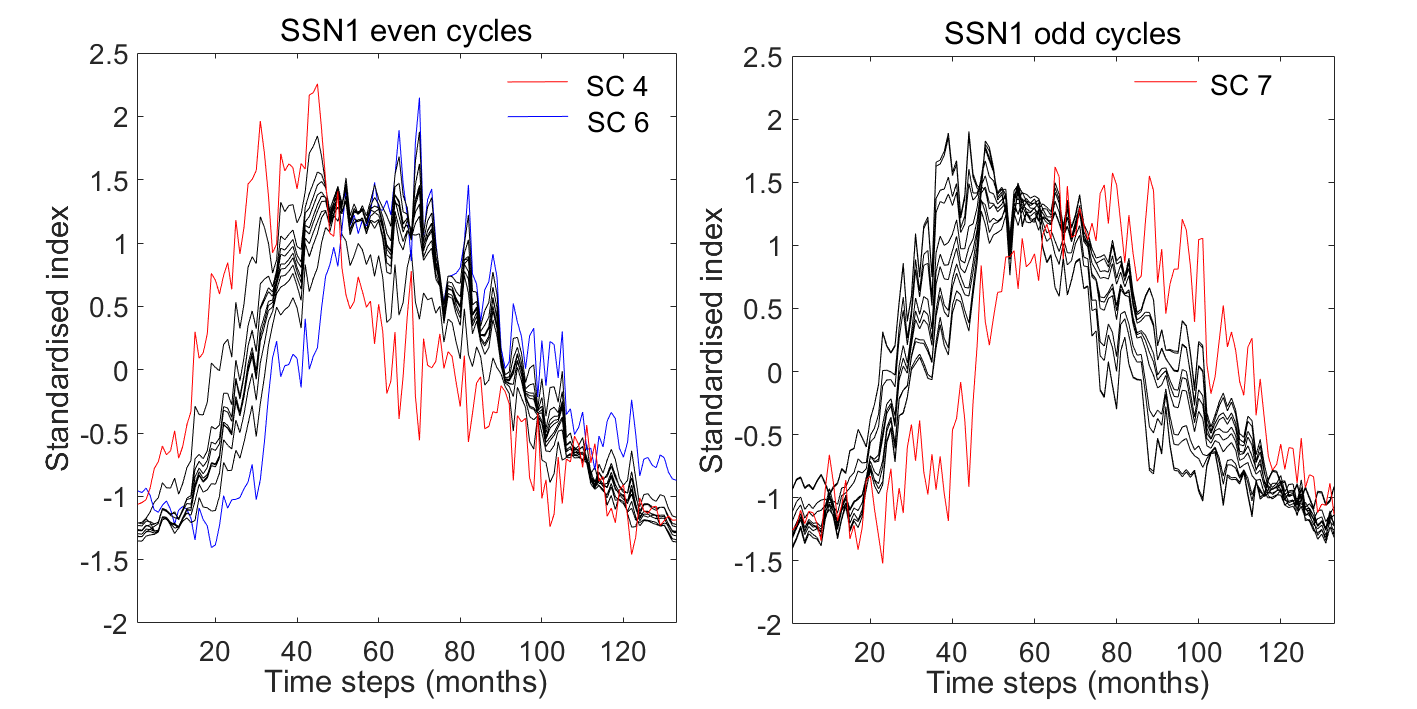}
                \caption{Scaled sums of PC1+PC2 for a) even SSN1 cycles and b) odd GSN cycles.}
                \label{fig:even_odd_PC1+PC2_curves}
\end{figure*}

Figures \ref{fig:even_odd_PC1+PC2_curves}a and \ref{fig:even_odd_PC1+PC2_curves}b show the scaled sums of PC1+PC2 of all SSN1 even and odd cycles, respectively. Even though the variation is quite strong elsewhere, especially in the odd cycles, the cycles are very similar to each other after the maximum in the region of the Gnevyshev gap in both cases. This suggests that the Gnevyshev gap is a common fundamental property of sunspot cycles that divides the sunspot cycle into two rather disparate parts: the ascending and maximum phase, and the declining phase \citep{Takalo_2018}. Moreover, the even cycles have a flat and wide maximum, while odd cycles have a single-peak maximum and the ascending phase starts slightly after this. The red (SC4) and blue curves (SC6) in panel \ref{fig:even_odd_PC1+PC2_curves}a and the red curve (SC7) in panel \ref{fig:even_odd_PC1+PC2_curves}b show the cycles that differ most from the other cycles.

\begin{figure*}
        \centering
        \includegraphics[width=0.8\textwidth]{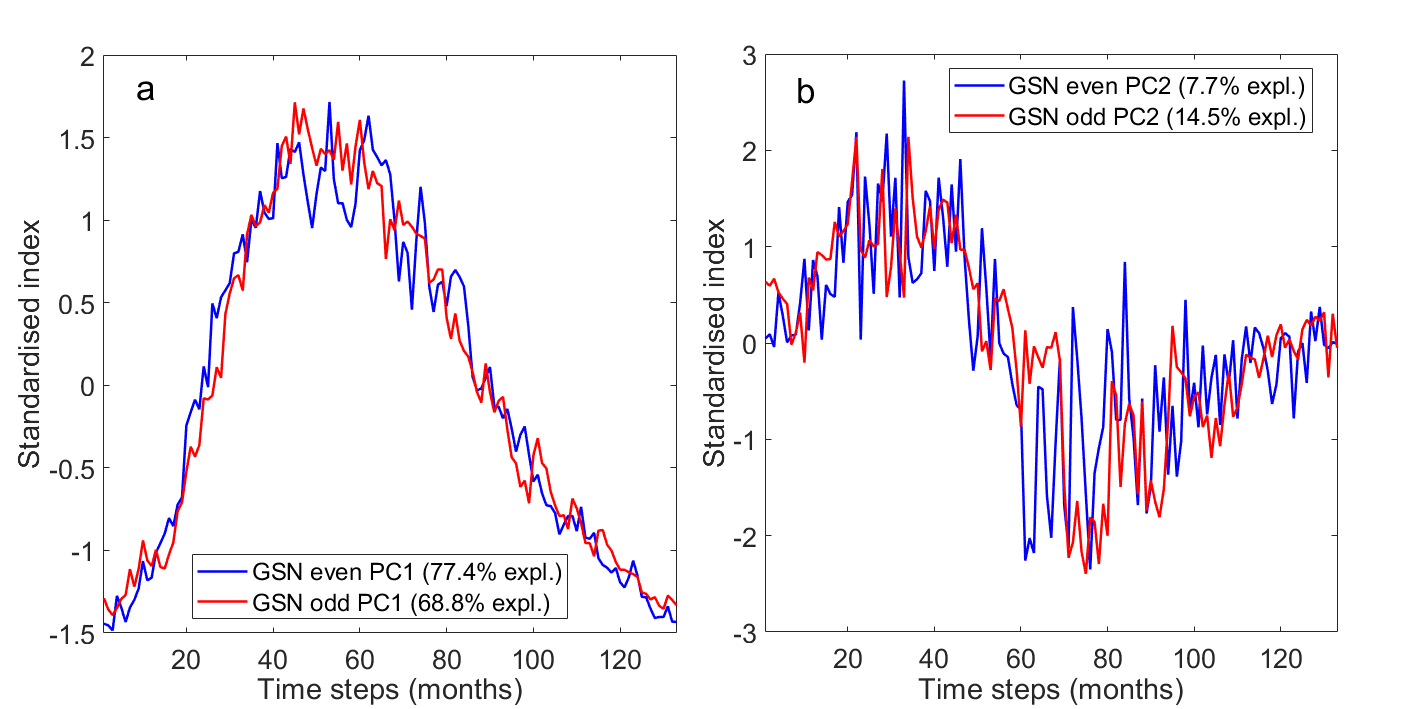}
                \caption{a) Principal components 1 and b) PC2s for GSN even and odd solar cycles 1-23.}
                \label{fig:GSN_even_odd_PCs}
\end{figure*}

\begin{figure*}
        \centering
        \includegraphics[width=0.8\textwidth]{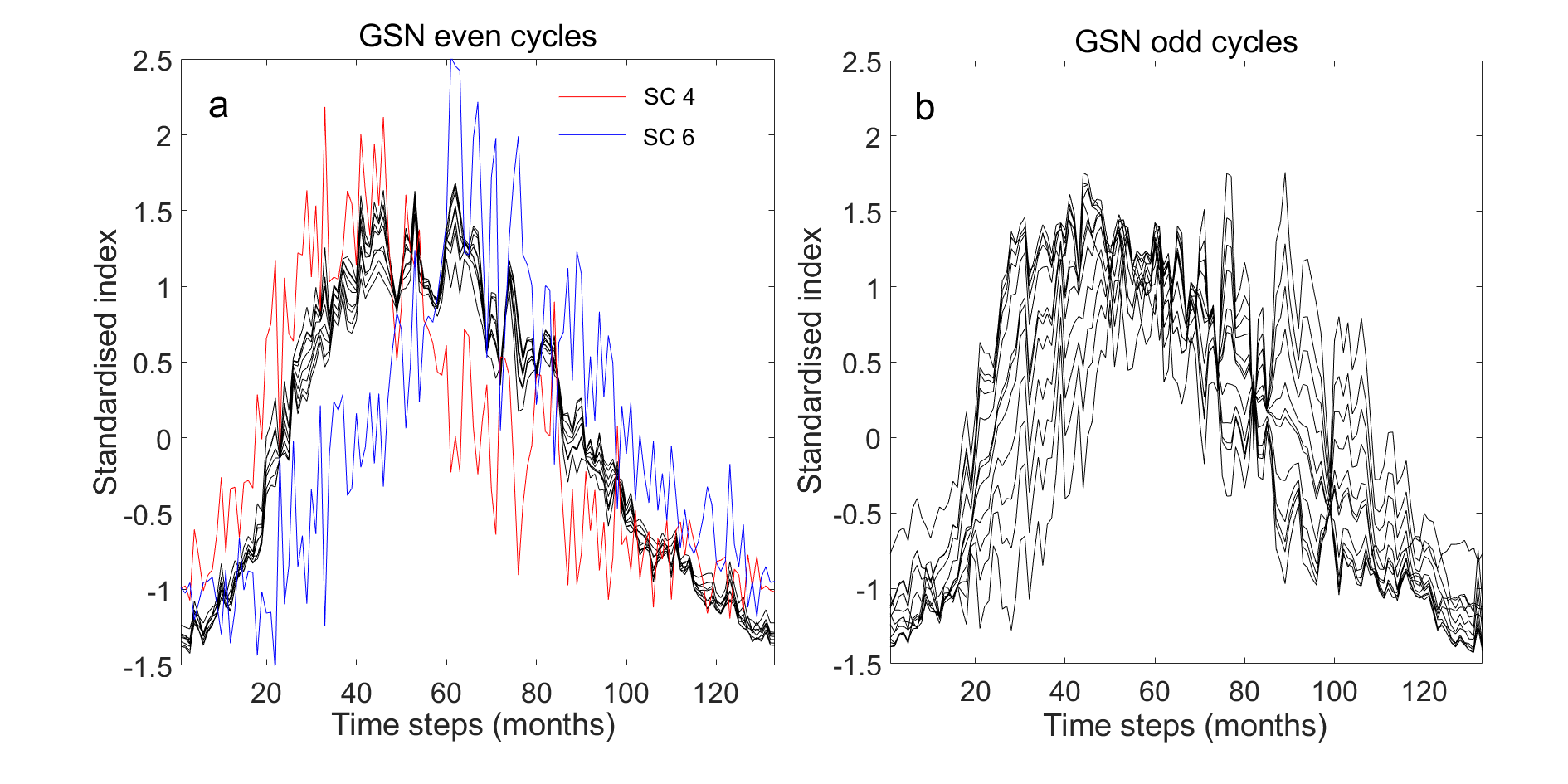}
                \caption{Scaled sums of PC1+PC2 for a) even GSN cycles and b) odd GSN cycles.}
                \label{fig:GSN_even_odd_PC1+PC2_curves}
\end{figure*}

We applied a similar PCA to even and odd GSN cycles separately. Figure \ref{fig:GSN_even_odd_PCs} shows the first and second  principal components of even and odd solar cycles in panels \ref{fig:GSN_even_odd_PCs}a and \ref{fig:GSN_even_odd_PCs}b, respectively. The PC1s explain 77.4\% and 68.8\%, and PC2s explain 7.7\% and 14.5\% of the total variance of even and odd cycles, respectively. The total variation thus explained by the first two PCs is 85.1\% for even and 83.3\% for odd cycles. The main difference, however, is that PC1 explains almost 9\% more for the even cycles than for odd cycles. The reason for this is shown in Fig. \ref{fig:GSN_even_odd_PC1+PC2_curves}, where we show the scaled sums of PC1+PC2 of all GSN even and odd cycles. Fig.\ref{fig:GSN_even_odd_PC1+PC2_curves}a shows that except for the two cycles SC4 (red curve) and SC6 (blue curve), the cycle curves are very similar to each other, while the cycle curves of Fig. \ref{fig:GSN_even_odd_PC1+PC2_curves}b for odd cycles have huge mutual variation. This may partly be due to variance in the length of the cycles. When we leave out the somewhat anomalously long cycles SC4 and SC6, the variances in length are 128.4 and 155.5 for even and odd cycles, respectively. When we leave SC4 and SC6 out of the PCA, the PC1 alone accounts for 84.5\% of the total variance for even cycles. We note, especially, that the GG is more distinct for even GSN cycles than for odd GSN cycles.

\begin{figure}
        \centering
        \includegraphics[width=0.5\textwidth]{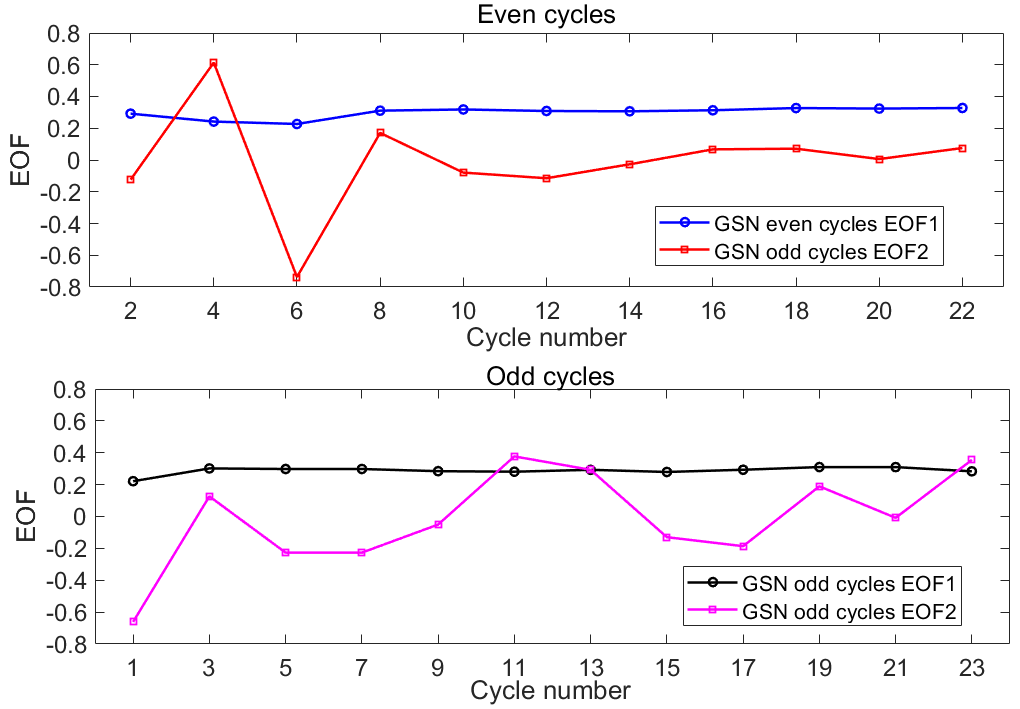}
                \caption{First two EOFs of a) even GSN and b) odd GSN cycles.}
                \label{fig:EOFs_even_odd_GSN} 
\end{figure}

\section{PCA of sunspot area data for even and odd cycles 12-23}

\begin{figure}
        \centering
        \includegraphics[width=0.5\textwidth]{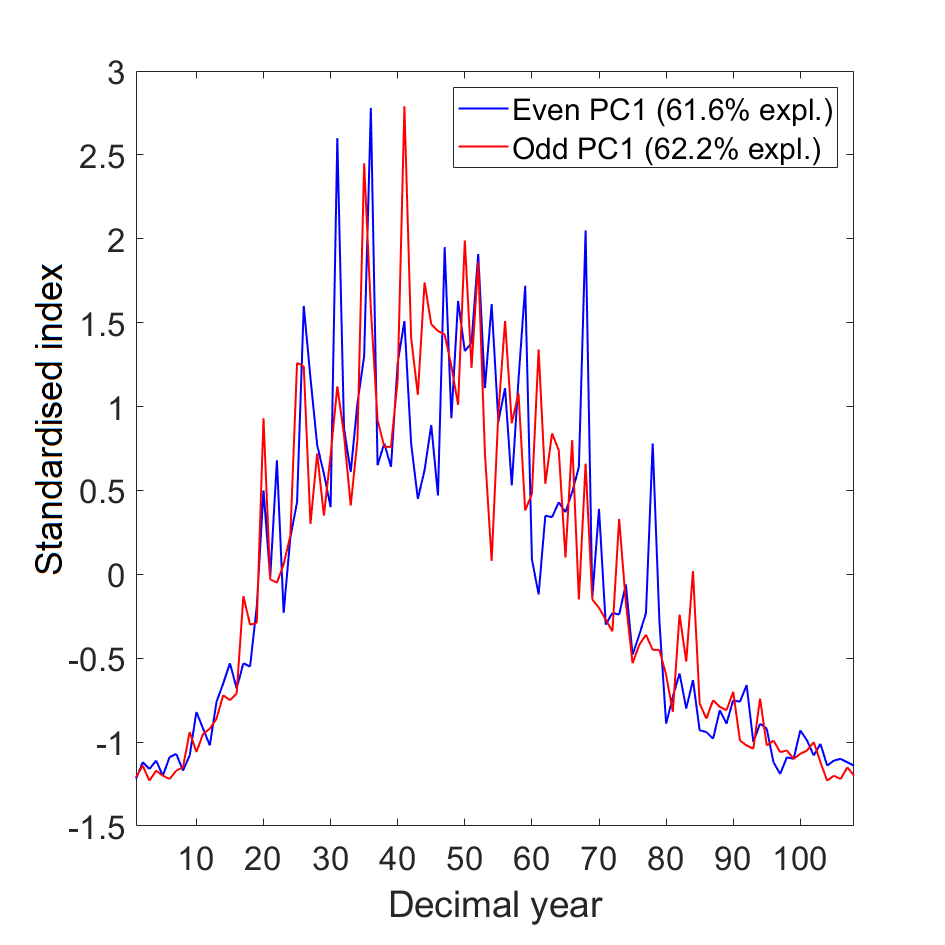}
                \caption{Principal components 1 of sunspot area data for even and odd solar cycles 12-23.}
                \label{fig:Even_odd SA_PCs}
\end{figure}

\begin{figure}
        \centering
        \includegraphics[width=0.5\textwidth]{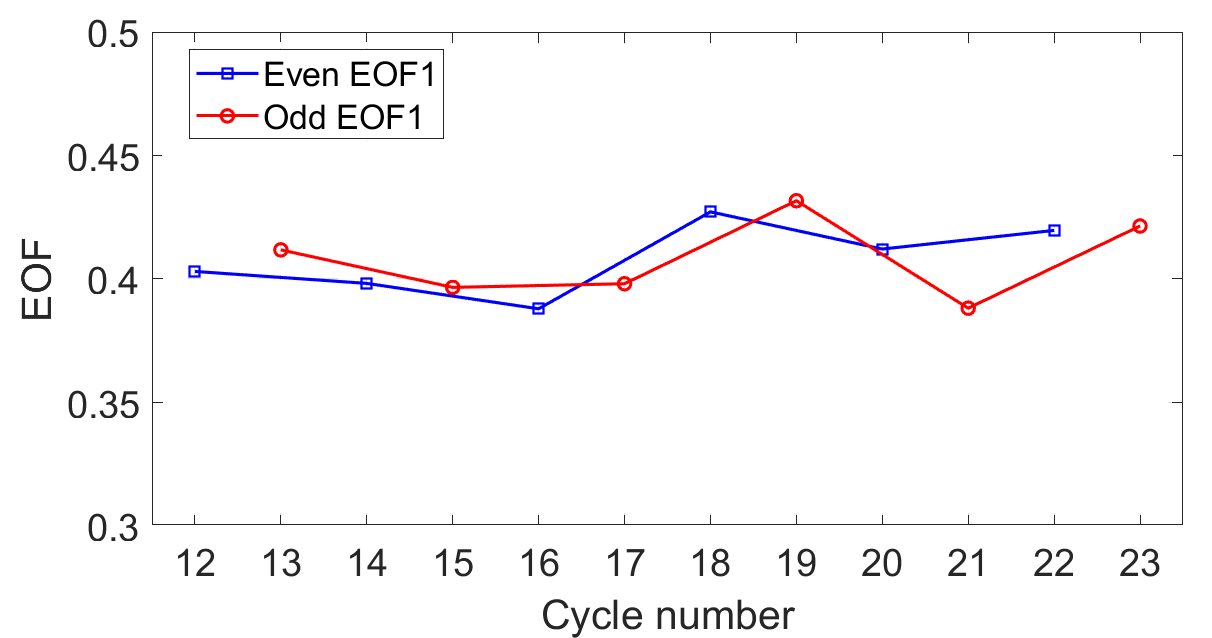}
                \caption{Empirical orthogonal function 1 of sunspot area data for even and odd solar cycles 12-23.}
                \label{fig:Even_odd SA_EOF1s}
\end{figure}

In studying the temporal distribution, we need to standardise the lengths of the cycles in some way. Because the time stamps in the database of the sunspot area data is expressed as decimal years and there are many simultaneous sunspots, we used a different standardising than before. We resampled all cycles such that their length was the average cycle length for SC12-SC23, that is, 10.8 years, and presented this as multiples of 0.1 year.
Figure \ref{fig:Even_odd SA_PCs} shows the leading principal component for even and odd solar cycles for the temporal evolution of the entire area in SC12-SC23. The PC1s explain 61.6\% and 62.2\% and of the total variance of even and odd cycles, respectively. The PC1s are more peaky than for the earlier SSN1 data, but the peaks seem to be (almost) in the same sites for even and odd data. The greatest difference is at 42-46 decimal years (4.2-4.6 years), where even cycles have a far smaller area than odd cycles. Figure \ref{fig:Even_odd SA_EOF1s} shows the first EOFs for the even and odd solar cycle sunspot area data. Although the EOF1 for all cycles is significant, cycles 18 and 19 have the greatest weight for even and odd PC1, respectively.   
The other PCs are quite noisy and carry information only on some individual cycles, therefore we do not show them here. Principal components 2-4 account for 13.8, 8.0 and 7.0 \% and 10.8, 9.5 and 7.5 \% for even and odd cycles, respectively.

\section{Temporal analysis of sunspot areas and sunspot groups for even and odd solar cycles}

\subsection{Temporal evolution of sunspot areas}

\begin{figure*}
        \centering
        \includegraphics[width=0.8\textwidth]{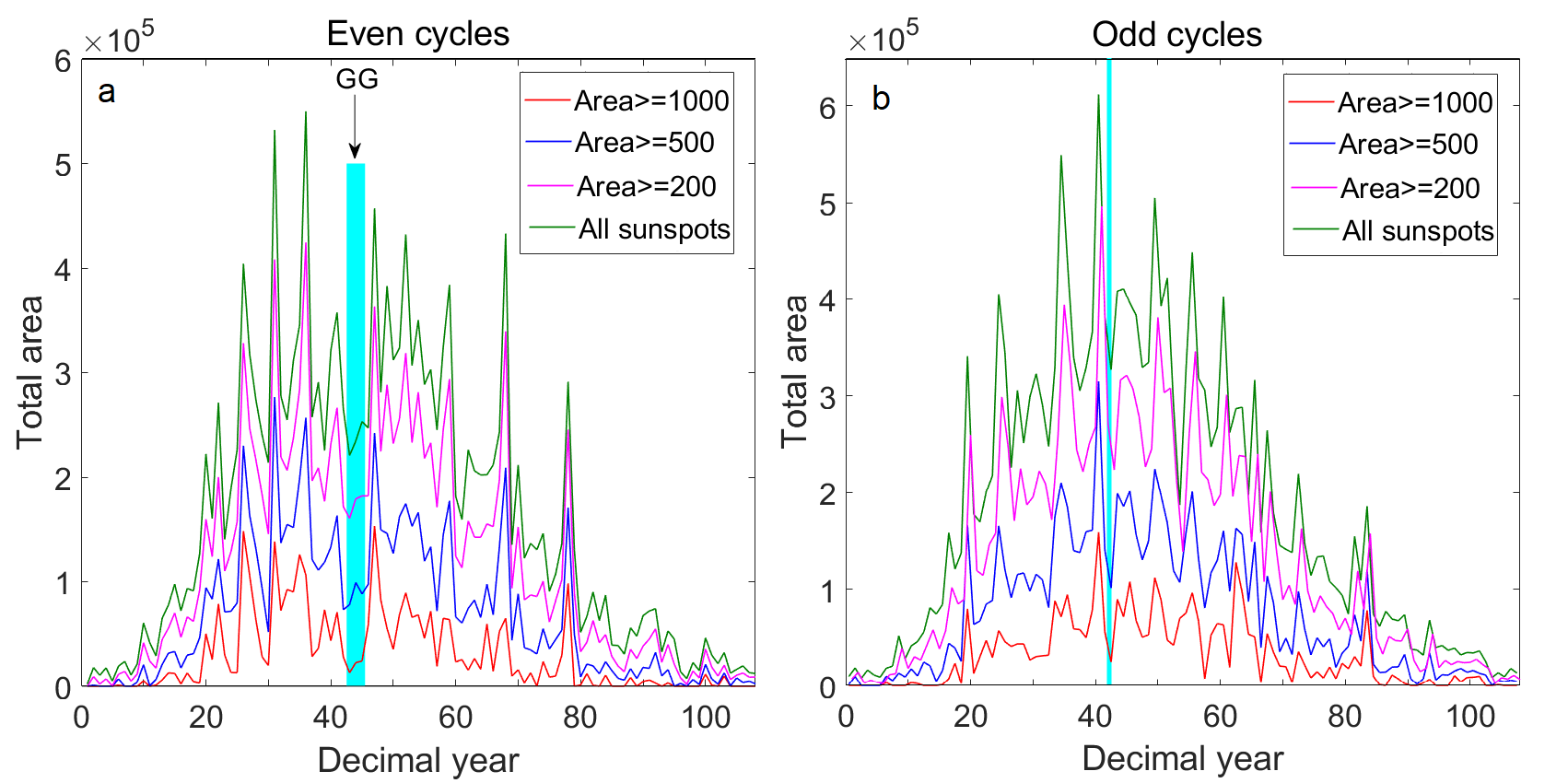}
                \caption{a) Total area for sunspots equal to or exceeding 1000 MH, 500 MH, and 200 MH and for all sunspots of even cycles as a function of decimal year (unit=0.1 year).b) Same as a), but for sunspots of odd cycles.}
                \label{fig:Even_odd_total_areas}
\end{figure*}

Figure \ref{fig:Even_odd_total_areas}a and \ref{fig:Even_odd_total_areas}b show the total area for sunspots equal to or exceeding 1000 MH, 500 MH, and 200 MH and for all sunspots, respectively. The GG is shown in Fig. \ref{fig:Even_odd_total_areas}a as a cyan bar, and it is seen even more clearly here in all of the aforementioned groups of sunspots. The two-sample T-test gives p-values for the unequal mean values for the interval 42-46 with p=0.015 (area>=1000) , p=0.0094 (area>=500), p=0.020 (area>=200), and p=0.037 (all sunspots) compared to the areas in the year before and the year after the gap. If we a priori assume that the GG interval might have a lower mean value, the p-values are half of the aforementioned p-values (one-sided T-test). In this way, the significance of the lower mean total area for the GG interval is at least at a level of about 95\% for all sunspots of even cycles. In addition to the smaller number of sunspots in this interval, they are smaller at the GG interval than in the surrounding sunspots. The average size of the sunspots in the interval 42-46 for even cycles is 152 MH, while the average area of the surrounding sunspots (a year before and a year after) is 188 MH. The T-test for the difference of the means is 0.0025 for the period 42-46 against one year before and one year after the period. The odd cycles (Fig. \ref{fig:Even_odd_total_areas}b) have only a small gap at 42-43 decimal year, and its is insignificant with p=0.22 (p=0.11 for one-sided T-test) compared against the null hypothesis with similar mean values for the one-year intervals before and after the gap. The average size of the sunspots in the small interval 42-43 for odd cycles is 153 MH, while the average area of the surrounding sunspots is 168 MH, but for the interval 42-46, it is the same size on average as for the surrounding sunspots.

\subsection{Temporal distribution of sunspot groups for even and odd cycles}

\begin{figure*}
        \centering
        \includegraphics[width=0.75\textwidth]{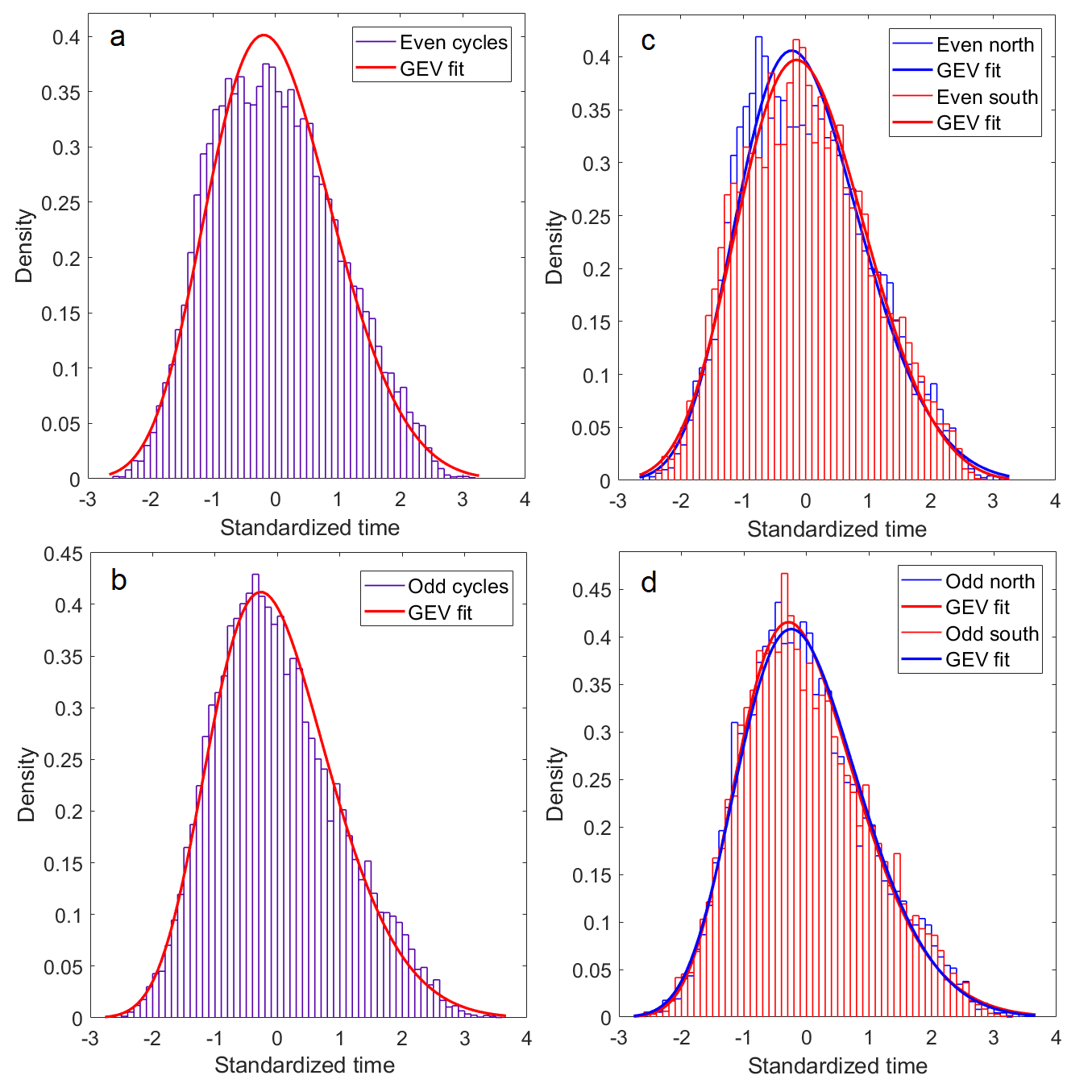}
                \caption{a) Standardised temporal distributions of sunspot groups for even cycles between SC8-SC23. b) Same as a), but for odd cycles. c) Standardised temporal distribution for the northern and southern sunspot groups of even cycles. d) Same as c), but for odd cycles.}
                \label{fig:Temporal_distribution_even_odd_SPs}
\end{figure*}

Because the length of the wings of the sunspot groups varies and they are not concurrent, we have to standardise the time axis. Moreover, because the wings of the sunspots are partly overlaid (see Fig. \ref{fig:Leussu_data}), we standardised them simply by calculating time as $x_{i}=\left(t_{i}-\overline{t}\right)/std(t)$, where $t_{i}$ is the original decimal year of each group, $\overline{t}$ is the mean time of the groups in each wing, and $std(t)$ is the standard deviation of the $t_{i}$s. Figures \ref{fig:Temporal_distribution_even_odd_SPs}a and \ref{fig:Temporal_distribution_even_odd_SPs}b show the standardised temporal distributions of sunspot groups for even and odd cycles between SC8-SC23, respectively. The negative log-likelihood (NLogL) of the generalised extreme value (GEV) distribution fits for even and odd wing sunspots is  33661 and 37176, while the NLogL for normal distribution fits is  34004 and 37787. The location ($\mu$, scale ($\sigma$) and shape (k) for the even and odd GEV fit (with standard errors) are -0.390 (0.0068) , 0.937 (0.0049), -0.199 (0.0049), and -0.410 (0.0062), 0.903 (0.0044), -0.147 (0.00448), respectively. The most distinctive difference between even and odd cycles is that the distribution of odd cycles is more leptokurtic and skewed more to the right than the distribution of even cycles. The skewnesses are 0.37 and 0.49 and the kurtoses are 2.79 and 2.94 for even and odd cycles, respectively.  Figure \ref{fig:Temporal_distribution_even_odd_SPs}c and \ref{fig:Temporal_distribution_even_odd_SPs}d show the same as panels \ref{fig:Temporal_distribution_even_odd_SPs}a and b, but for the distributions of the northern and southern sunspot groups separately. Figure \ref{fig:Temporal_distribution_even_odd_SPs}c shows that the double peak arises partly because the peak of the northern groups occurs earlier than the peak of the southern groups. This is because the distribution of the even northern sunspot groups is far more skewed to the right (positively) than the distribution of the even south sunspot groups, that is, the skewnesses are 0.43 and 0.31 for the northern and southern groups, respectively. The trough between them is at about one-third of the total standardised time of the distribution. This is probably the Gnevyshev gap, which is located approximately 33-42\% after the start of an individual cycle \citep{Takalo_2018}. The two-sample T-test shows that the gap is significant at the 95\% level with a p-value of 0.026. The skewnesses of odd cycles are 0.52 and 0.46 for the northern and southern groups, respectively.

\begin{figure*}
        \centering
        \includegraphics[width=0.8\textwidth]{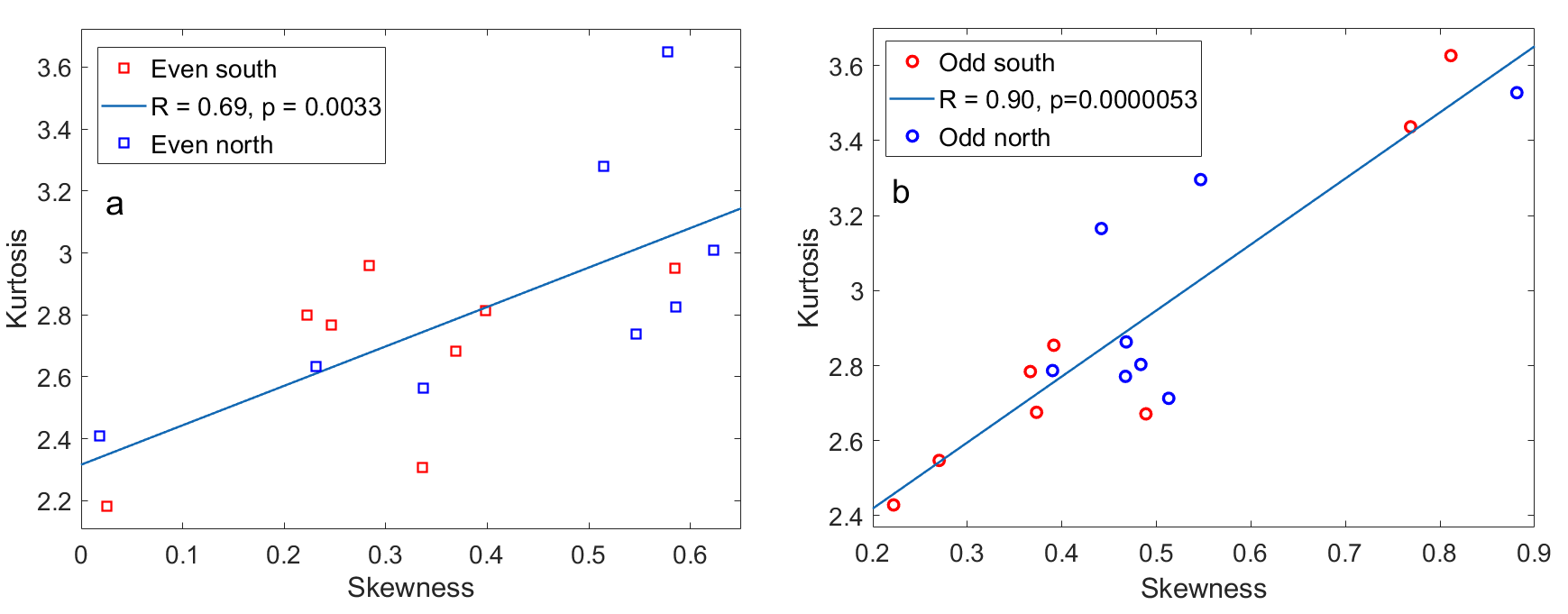}
                \caption{a) Skewness and kurtosis regression analysis for even cycles. b) Same as a), but for odd cycles.}
                \label{fig:regression_analysis}
\end{figure*}

Because of the differences in the skewness (third central moment) and kurtosis (fourth central moment) of the even and odd cycle sunspot groups, we studied their kurtoses as a function of skewness separately. Figure \ref{fig:regression_analysis} shows the skewness-kurtosis plane for even southern (red squares) and northern (blue squares) cycles and odd southern (red circles) and northern (blue circles) cycles in panels a and b, respectively. There is a significant correlation between skewness and kurtosis (R=0.69, p=0.0033) for even cycles and a still better correlation (R=0.90, p=0.0000053) for odd cycles. This resembles the Waldmeier rule, that is, that the ascending phase length and cycle height are anti-correlated. However, according to our studies, the (anti-)correlation of the Waldmeier rule for all even GSN cycles between SC1-SC23 is -0.715 (p=0.013), which is significant, but for all odd cycles it is only -0.242 (p=0.45) and thus insignificant. The sunspot group data are different than the GSN, and the kurtosis does not mean that a cycle is high, therefore these result are not as such comparable.

\section{Conclusions}

We have studied the Z\"urich sunspot number series and the group sunspot number series for sunspot cycles 1-23 using the principal component analysis separately for even and odd cycles. We used the standard cycle minima and lengths for the SSN1 data \citep{NGDC_2013}, but calculated the minima and lengths for the GSN using the 13-month Gleissberg filter. We resampled the monthly sunspot values such that all cycles have the same length of 133 months. Before applying the PCA, we standardised each individual cycle to have zero mean and unit standard deviation \citep{Takalo_2018}. In this way, the cycle amplitudes do not affect their common shape. The first two components of the analysis explain 77.2\% and 79.6\%  of the total variance for even and odd cycles of SSN, respectively, and 77.4\% and 68.8\% of the total variance for even and odd cycles of GSN, respectively. PC1 describes the average shape of the solar cycle (the "model" cycle), and PC2 represents the leading correction of individual cycles from the model cycle \citep{Takalo_2018}.

We found that the shape of even cycles is more homogeneous than the shape of odd cycles. The variation in the shape of the odd cycles in the declining part of the cycle is huge, especially for GSN data. The analysis also suggests that we have too few and/or inaccurate measurements during the early cycles before SC8. Even cycles are more double peaked than odd cycles, which seem to have only one clear peak and a small gap after it, but no clear other peak, but the descending phase starts gradually after the gap.

The temporal evolution of sunspot areas for even cycles shows a lack of large sunspots after four years (exactly between 42-46 decimal years), that is, at about 40\% after the start of the cycle. This gap is first seen in the PCA of sunspot area data and is then better visible in the analysis of different size area data. This is related to the Gnevyshev gap and is consistent with the earlier result by \cite{Takalo_2018}. The significance level of this gap for even cycles is at least 95\% for all sunspots. Furthermore, the average size of the sunspots is smaller in this gap than one year before or one year after the gap. For odd cycles the gap is narrower (42-43 decimal years), and it is insignificant according to the two-sample T-test for all sunspots and large sunspots.

The sunspot group distribution analysis shows that the most distinctive difference between even and odd cycles is that the distribution of odd cycles is more leptokurtic and skewed more to the right than the distribution of even cycles. The skewnesses are 0.37 and 0.49 and kurtoses 2.79 and 2.94 for even and odd cycles, respectively. We also find that the distribution of even cycles has  a double-peak structure, which arises partly because the peak of the northern groups occurs earlier than the peak of the southern groups. This is because the distribution of even northern sunspots groups is much more skewed to the right (positively) than the distribution of even southern sunspot groups, that is, the skewnesses are 0.43 and 0.31 for the northern and southern groups, respectively.

We also present another Waldmeier-type rule, that is, we find a correlation between skewness and kurtosis of the sunspot group cycles. The correlation coefficient for even cycles is 0.69, and for odd cycles, it is 0.90. The overall correlation (both even and odd cycles) is R=.72 (p = $3.7\times10^{-6}$). 

\begin{acknowledgements}

We acknowledge the financial support by the Academy of Finland to the ReSoLVE Centre of Excellence (project no. 272157). The sunspot data were obtained from WDC-SILSO, Royal Observatory of Belgium, Brussels (http://sidc.be/silso/) and the sunspot area data from RGO-USAF/NOAA \newline (https://solarscience.msfc.nasa.gov/greenwch.shtml). The dates of the cycle minima and the lengths of the SSN cycles were obtained from from the National Geophysical Data Center (NGDC), Boulder, Colorado, USA (ftp.ngdc.noaa.gov). 

\end{acknowledgements}

\bibliographystyle{aa}

%\bibliography{K:/TeX/references}
\bibliography{references_JT}

\begin{thebibliography}{45}
\expandafter\ifx\csname natexlab\endcsname\relax\def\natexlab#1{#1}\fi

\bibitem[{{Ahluwalia} \& {Kamide}(2004)}]{Ahluwalia_2004}
{Ahluwalia}, H.~S. \& {Kamide}, Y. 2004, in COSPAR Meeting, Vol.~35, 35th
  COSPAR Scientific Assembly, ed. J.-P. {Paill{\'e}}, 470

\bibitem[{{Badalyan} \& {Obridko}(2017)}]{Badalyan_2017}
{Badalyan}, O.~G. \& {Obridko}, V.~N. 2017, Astron. Astrophys., 603

\bibitem[{{Bazilevskaya} {et~al.}(2006){Bazilevskaya}, {Makhmutov}, \&
  {Sladkova}}]{Bazilevskaya_2006}
{Bazilevskaya}, G.~A., {Makhmutov}, V.~S., \& {Sladkova}, A.~I. 2006, Adv.\
  Space\ Res., 38, 484

\bibitem[{{Carbonell} {et~al.}(2007){Carbonell}, {Terradas}, {Oliver}, \&
  {Ballester}}]{Carbonell_2007}
{Carbonell}, M., {Terradas}, J., {Oliver}, R., \& {Ballester}, J.~L. 2007,
  Astron. Astrophys., 476, 951

\bibitem[{{Chang}(2012)}]{Chang_2012}
{Chang}, H.-Y. 2012, \na, 17, 247

\bibitem[{{Chatzistergos} {et~al.}(2017){Chatzistergos}, {Usoskin},
  {Kovaltsov}, {Krivova}, \& {Solanki}}]{Chatzistergos_2017}
{Chatzistergos}, T., {Usoskin}, I.~G., {Kovaltsov}, G.~A., {Krivova}, N.~A., \&
  {Solanki}, S.~K. 2017, Astron. Astrophys., 602, 18

\bibitem[{{Chowdhury} {et~al.}(2019){Chowdhury}, {Kilcik}, {Yurchyshyn},
  {Obridko}, \& {Rozelot}}]{Chowdhury_2019}
{Chowdhury}, P., {Kilcik}, A., {Yurchyshyn}, V., {Obridko}, V.~N., \&
  {Rozelot}, J.~P. 2019, Sol.\ Phys., 294, 142

\bibitem[{{Clette} {et~al.}(2014){Clette}, {Svalgaard}, {Vaquero}, \&
  {Cliver}}]{Clette_2014}
{Clette}, F., {Svalgaard}, L., {Vaquero}, J.~M., \& {Cliver}, E.~W. 2014,
  Space\ Sci.\ Rev., 186, 35

\bibitem[{{Coles}(2001)}]{Coles_2001}
{Coles}, S. 2001, {An Introduction to Statistical Modeling of Extreme Values}
  (Springer-Verlag London Limited 2001)

\bibitem[{{Derrick} {et~al.}(2016){Derrick}, {Deirdre}, \&
  {White}}]{Derrick_2016}
{Derrick}, B., {Deirdre}, T., \& {White}, P. 2016, The Quantitative Methods for
  Psychology, 12, 30

\bibitem[{{Du}(2015)}]{Du_2015}
{Du}, Z.~L. 2015, Astrophys.\ J., 804, 15

\bibitem[{{Feminella} \& {Storini}(1997)}]{Feminella_1997}
{Feminella}, F. \& {Storini}, M. 1997, Astron. Astrophys., 322, 311

\bibitem[{{Forbes} {et~al.}(2011){Forbes}, {Evans}, {Hastings}, \&
  {Peacock}}]{Forbes_2011}
{Forbes}, C., {Evans}, N., {Hastings}, N., \& {Peacock}, B. 2011, {Statistical
  Distributions} (John Wiley Sons, Inc., Hoboken, New Jersey.), 47--49

\bibitem[{{Gnevyshev}(1967)}]{Gnevyshev_1967}
{Gnevyshev}, M.~N. 1967, Sol.\ Phys., 1, 107

\bibitem[{{Gnevyshev}(1977)}]{Gnevyshev_1977}
{Gnevyshev}, M.~N. 1977, Sol.\ Phys., 51, 175

\bibitem[{{Hathaway}(2015)}]{Hathaway_2015}
{Hathaway}, D.~H. 2015, Living\ Rev.\ Solar\ Phys., 12, 4

\bibitem[{{Hoyt} \& {Schatten}(1998)}]{Hoyt_1998}
{Hoyt}, D.~V. \& {Schatten}, K.~H. 1998, Sol.\ Phys., 179, 189

\bibitem[{{Ivanov} {et~al.}(2011){Ivanov}, {Miletskii}, \&
  {Nagovitsyn}}]{Ivanov_2011}
{Ivanov}, V.~G., {Miletskii}, E.~V., \& {Nagovitsyn}, Y.~A. 2011, Astron.Rep.,
  55, 911

\bibitem[{{Javaraiah}(2012)}]{Javaraiah_2012}
{Javaraiah}, J. 2012, Sol.\ Phys., 281, 827

\bibitem[{{Javaraiah}(2016)}]{Javaraiah_2016}
{Javaraiah}, J. 2016, Astrophys. Space Sci., 361, 208

\bibitem[{{Jiang} {et~al.}(2011){Jiang}, {Cameron}, {Schmitt}, \&
  {Sch{\"u}ssler}}]{Jiang_2011}
{Jiang}, J., {Cameron}, R.~H., {Schmitt}, D., \& {Sch{\"u}ssler}, M. 2011,
  Astron. Astrophys., 528, A82

\bibitem[{{Jolliffe}(2002)}]{Jolliffe_2002}
{Jolliffe}, I.~T. 2002, {Principal component analysis} (Springer-Verlag, New
  York, 2nd ed.)

\bibitem[{{Jolliffe} \& {Cadima}(2016)}]{Jolliffe_2016}
{Jolliffe}, I.~T. \& {Cadima}, J. 2016, Philosophical Transactions of the Royal
  Society of London Series A, 374, 20150202

\bibitem[{{Kotz} \& {Nadajarah}(2000)}]{Kotz_2000}
{Kotz}, S. \& {Nadajarah}, S. 2000, {Extreme Value Distributions: Theory and
  Applications.} (London: Imperial College Press.)

\bibitem[{{Krishnamoorthy}(2006)}]{Krishnamoorthy_2006}
{Krishnamoorthy}, K. 2006, Handbook of Statistical Distributions with
  Applications (Chapman \& Hall/CRC, Taylor \& Francis Group, Boca Raton, FL
  33487-2742)

\bibitem[{{Leussu} {et~al.}(2016{\natexlab{a}}){Leussu}, {Usoskin}, {Arlt}, \&
  {Mursula}}]{Leussu_2016}
{Leussu}, R., {Usoskin}, I.~G., {Arlt}, R., \& {Mursula}, K.
  2016{\natexlab{a}}, Astron. Astrophys., 592, A160

\bibitem[{{Leussu} {et~al.}(2016{\natexlab{b}}){Leussu}, {Usoskin}, {Senthamizh
  Pavai}, {Diercke}, {Arlt}, \& {Mursula}}]{Leussu_2017}
{Leussu}, R., {Usoskin}, I.~G., {Senthamizh Pavai}, V., {et~al.}
  2016{\natexlab{b}}, VizieR Online Data Catalog, 359

\bibitem[{{Li} {et~al.}(2009){Li}, {Gao}, \& {Zhan}}]{Li_2009}
{Li}, K.~J., {Gao}, P.~X., \& {Zhan}, L.~S. 2009, Sol.\ Phys., 254, 145

\bibitem[{{Mandal} {et~al.}(2017){Mandal}, Bidya Binay~{Karak}, \&
  D.}]{Mandal_2017}
{Mandal}, S., Bidya Binay~{Karak}, B.~B., \& D., B. 2017, Astrophys.\ J., 851,
  70

\bibitem[{{Munoz-Jaramillo} {et~al.}(2015){Munoz-Jaramillo}, {Senkpeil},
  {Windmueller}, {Amouzou}, {Longcope}, {Tlatov}, {Nagovitsyn}, Alexei
  A.~{Pevtsov}, {Chapman}, {Cookson}, {Yeates}, {Watson}, {Balmaceda},
  {DeLuca}, \& {Martens}}]{Munoz-Jaramillo_2015}
{Munoz-Jaramillo}, A., {Senkpeil}, R.~R., {Windmueller}, J.~C., {et~al.} 2015,
  Astrophys.\ J., 800, 48

\bibitem[{{Mursula} {et~al.}(2001){Mursula}, {Usoskin}, \&
  {Kovaltsov}}]{Mursula_2001}
{Mursula}, K., {Usoskin}, I.~G., \& {Kovaltsov}, G.~A. 2001, \solphys, 198, 51

\bibitem[{{NGDC}(2013)}]{NGDC_2013}
{NGDC}. 2013, {Solar-indices}, the data via anonymous FTP
  (\texttt{ftp.ngdc.noaa.gov}) from the National Geophysical Data Center
  (NGDC), Boulder, Colorado, USA

\bibitem[{{Norton} \& {Gallagher}(2010)}]{Norton_2010}
{Norton}, A.~A. \& {Gallagher}, J.~C. 2010, Sol.\ Phys., 261, 193

\bibitem[{RGO-USAF/NOAA(2017)}]{Nasa_MSFC_2017}
RGO-USAF/NOAA. 2017, https://solarscience.msfc.nasa.gov/greenwch.shtml

\bibitem[{{Santos} {et~al.}(2015){Santos}, {Cunha}, {Avelino}, \&
  {Campante}}]{Santos_2015}
{Santos}, A.~R.~G., {Cunha}, M.~S., {Avelino}, P.~P., \& {Campante}, T.~L.
  2015, Astron. Astrophys., 580, A62

\bibitem[{{Snedecor} \& {Cochran}(1989)}]{Snedecor_1989}
{Snedecor}, G.~W. \& {Cochran}, W.~G. 1989, {Statistical Methods} (Eighth
  Edition, Iowa State University Press)

\bibitem[{{Storini} {et~al.}(2003){Storini}, {Bazilevskaya}, {Fluckiger},
  {Krainev}, {Makhmutov}, \& {Sladkova}}]{Storini_2003}
{Storini}, M., {Bazilevskaya}, G.~A., {Fluckiger}, E.~O., {et~al.} 2003, Adv.\
  Space\ Res., 31, 895

\bibitem[{{Takalo}(2020)}]{Takalo_2020}
{Takalo}, J. 2020, Solar Physics, accepted

\bibitem[{{Takalo} \& {Mursula}(2018)}]{Takalo_2018}
{Takalo}, J. \& {Mursula}, K. 2018, Astron. Astrophys., 620

\bibitem[{{Temmer} {et~al.}(2006){Temmer}, {Ryb{\'a}k}, {Bend{\'\i}k},
  {Veronig}, {Vogler}, {Otruba}, {P{\"o}tzi}, \& {Hanslmeier}}]{Temmer_2006}
{Temmer}, M., {Ryb{\'a}k}, J., {Bend{\'\i}k}, P., {et~al.} 2006, Astron.
  Astrophys., 447, 735

\bibitem[{{Vernova} {et~al.}(2016){Vernova}, {Tyasto}, \&
  {Baranov}}]{Vernova_2016}
{Vernova}, E.~S., {Tyasto}, M.~I., \& {Baranov}, D.~G. 2016, Sol.\ Phys., 291,
  741

\bibitem[{{Waldmeier}(1935)}]{Waldmeier_1935}
{Waldmeier}, M. 1935, Astron.\ Mitt.\ Zurich, 14, 105

\bibitem[{{Waldmeier}(1939)}]{Waldmeier_1939}
{Waldmeier}, M. 1939, Astron.\ Mitt.\ Zurich, 14, 470

\bibitem[{{Zhang} {et~al.}(2018){Zhang}, {Li}, \& {Feng}}]{Zhang_2018}
{Zhang}, J., {Li}, F.-Y., \& {Feng}, W. 2018, Research in Astronomy and
  Astrophysics, 18, 007

\bibitem[{{Zharkov} {et~al.}(2005){Zharkov}, {Zharkova}, \&
  {Ipson}}]{Zharkov_2005}
{Zharkov}, S.~I., {Zharkova}, V.~V., \& {Ipson}, S.~S. 2005, Sol.\ Phys., 228,
  377–397

\end{thebibliography}

\section{Appendix A: PCA method}

Standardised sunspot cycles are collected into the columns of the data matrix X, which can be decomposed as
\begin{equation}
        X = U\:D\;V^{T}  \     ,
\end{equation}
where U and V are orthogonal matrices and D is a diagonal matrix 
        $D= diag\left(\lambda_{1},\lambda_{2},...,\lambda_{n}\right),$
with $\lambda_{i}$ denoting the $i\text{-th}$ singular value of matrix X in order of decreasing importance. The principal components are the column vectors of
\begin{equation}
  P = U\!D.
\end{equation}
The column vectors of the matrix V are called empirical orthogonal functions (EOF) and represent the weights of each principal component in the decomposition of each (standardised) cycle $X_{i}$, which can be approximated as 
\begin{equation}
        X_{i} = \sum^{N}_{j=1} \:P_{ij}\:V_{ij} \  ,
\end{equation}
where N is the number of principal components (here N=2). The variance explained by each PC is proportional to the square of the corresponding singular value $\lambda_{i}$. Hence the $i\text{-th}$
PC explains a percentage
\begin{equation}
\frac{\lambda^{2}_{i}}{\sum^{n}_{k=1}\!\lambda^{2}_{k}} \cdot\:100\%
\end{equation}
of the variance in the data.

\end{document}